\documentclass[12pt]{article}
\usepackage{graphicx}
\newskip\humongous \humongous=0pt plus 1000pt minus 100pt

\newif\ifdtup

\newcounter{eqnumber}[section]

\def\eqn#1{eq.~{(\ref{#1})}}

\def\fig#1{fig.~{\ref{#1}}}
\def\Fig#1{Figure~{\ref{#1}}}

\def\eps{\epsilon}
\def\ve{\varepsilon}
\def\M{{\cal M}}
\def\ds{\displaystyle}
\begin{document}
\bibliographystyle{unsrt}

\thispagestyle{empty}
\renewcommand{\thefootnote}{\fnsymbol{footnote}}
\begin{flushright}
{\small
SLAC--PUB--8435\\
hep-ph/0004143\\
April, 2000\\}
\end{flushright}

\vspace{.8cm}

\begin{center}
\begin{Large}
{\bf Recalculation of Proton Compton Scattering in Perturbative QCD
\footnote{Research supported by Department of Energy contract 
DE--AC03--76SF00515}}
\end{Large}

\vspace{1.5cm}

{T. Brooks\footnote{Supported in part by a National Science Foundation
graduate fellowship} and L. Dixon\\
\it Stanford Linear Accelerator Center\\
Stanford University\\
Stanford, CA 94309}

\medskip
\end{center}

\vfill

\begin{center}
{\bf\large   
Abstract }
\end{center}

\begin{quote}
At very high energy and wide angles, Compton scattering on the proton
($\gamma p \to \gamma p$) is described by perturbative QCD.  The
perturbative QCD calculation has been performed several times previously,
at leading twist and at leading order in $\alpha_s$, with mutually
inconsistent results, even when the same light-cone distribution
amplitudes have been employed.  We have recalculated the helicity
amplitudes for this process, using contour deformations to evaluate the
singular integrals over the light-cone momentum fractions.  We do not
obtain complete agreement with any previous result.  Our results are
closest to those of the most recent previous computation, differing
significantly for just one of the three independent helicity
amplitudes, and only for backward scattering angles.  We present results
for the unpolarized cross section, and for three different polarization
asymmetries.  We compare the perturbative QCD predictions for these
observables with those of the handbag and diquark models.
In order to reduce uncertainties associated with $\alpha_s$ and
the three-quark wave function normalization, we have normalized the 
Compton cross section using the proton elastic form factor.  The 
theoretical predictions for this ratio are about an order of magnitude
below existing experimental data.
\end{quote}

\vskip 2 cm

\begin{center} 
{\sl Submitted to Physical Review D}
\end{center}
\newpage

\pagestyle{plain}


\section{Introduction}

Exclusive real Compton scattering on the proton, $\gamma p \to \gamma p$,
is a promising arena for studying the short-distance structure of the
proton.  In the limit of large energy $\sqrt{s}$ and fixed scattering
angle $\theta$ in the center-of-mass frame, the real Compton amplitude
should factorize as the convolution of a perturbative hard scattering
matrix element with a nonperturbative light-cone distribution
amplitude~\cite{LB}.  The distribution amplitude is for the three valence
quarks in the proton; it describes how their longitudinal momentum is
partitioned when their transverse separation is very small.  Contributions
of Fock space states with more partons in the proton's light-cone wave
function should be suppressed by additional powers of $s$.  However, the
energy at which this asymptotic prediction of perturbative QCD (PQCD)
becomes valid is not known {\it a priori}.  Soft mechanisms such as the
soft overlap (or handbag) model~\cite{Rady,Diehl1,Diehl2} and the diquark
model~\cite{DiQuark,BergerThesis} could be comparable to, or even dominant 
over, the PQCD mechanism at the presently accessible center-of-mass
energies of 2--4 GeV.

The PQCD prediction for $\gamma p \to \gamma p$ contains a number of
uncertainties.  First, only the Born level has been computed;
next-to-leading-order corrections are likely to be large.  Related to
this, the Born level prediction is proportional to a high power of the
running strong coupling constant, $[ \alpha_s(\mu) ]^4$, and its
renormalization-scale ($\mu$) dependence leads to a large normalization
uncertainty on the cross section.  Second, the form of the proton
distribution amplitude is not well understood. Several groups have
produced model distribution amplitudes based primarily on QCD sum rule
analyses~\cite{CZ,GS,KS,COZ,SB,SBII,EHG}.  These distribution amplitudes
can lead to quite different predictions for the Compton helicity
amplitudes.  Most of the proposed distribution amplitudes tend to peak in a
region where two of the three quarks carry relatively small fractions $x$
of the proton longitudinal momentum.  This has led to skepticism about the
applicability of PQCD at accessible
energies~\cite{IsgurLS,BolzKroll,Rady,Diehl1}, because relatively soft
sub-processes (relative to $\sqrt{s}$) can reorient quarks with small $x$
from the initial proton direction to the final proton direction.

Despite all these caveats, it is still useful to know the PQCD predictions
for $\gamma p \to \gamma p$, if nothing else as an asymptotic limit.
There have already been four separate calculations at Born
level~\cite{MF,FZ,KN,GV}.  However, no two results agree with each other,
even when the same proton distribution amplitudes are assumed.  Given this
discrepancy in the literature, and the need for consistent predictions
from the PQCD mechanism, we undertook an independent recalculation of this
process.  Our results in fact differ from all previous work, although we
find reasonable agreement with ref.~\cite{KN} for a subset of the helicity
amplitudes, and excellent agreement with ref.~\cite{GV} for forward
scattering angles.

Our results are timely in view of the experimental situation.  For over
twenty years, the highest energy wide-angle Compton data available have
been from an experiment at Cornell~\cite{Shupe} which investigated the
energy range $4.6$~GeV$^2 < s < 12.1$~GeV$^2$.  These data appear to obey
an approximate $d\sigma/dt \propto s^{-6}$ scaling law, as predicted by PQCD,
although more precise data would be useful to confirm or refute this
behavior.  An experiment now underway at Jefferson Lab~\cite{JLab} should
soon improve the errors on the unpolarized cross section and its $\theta$-
and $s$-dependence, in the same kinematic range as the Cornell experiment.
This experiment also plans to measure a polarization asymmetry, the
transfer of longitudinal polarization from the incoming photon to the
outgoing proton, for at least one angle.  We shall discuss this asymmetry
further in section~\ref{ResultSection}.  An upgrade of the Jefferson Lab
electron beam to 12 GeV~\cite{JLabUpgrade} would allow for the very
important extension of this experiment to higher energies.  The proposed
ELFE facility~\cite{ELFE} with a 25 GeV electron beam would also be a
natural place to perform higher energy Compton measurements.

The remainder of this paper is organized as follows.  
In section~\ref{CalculationSection} we outline the calculation.  
In section~\ref{ResultSection} we present results for the unpolarized 
cross section and for some different polarization asymmetries.
Section~\ref{ConclusionSection} contains our conclusions.


\section{Calculation}
\label{CalculationSection}

Since the general PQCD calculational framework for the Compton process
has been described previously, e.g. in ref.~\cite{KN}, we will be brief
here.  The leading-twist PQCD factorization of the helicity amplitude
$\M^{\lambda\lambda'}_{hh'}$ for incoming (outgoing) photon helicity 
$\lambda$ ($\lambda'$) and proton helicity $h$ ($h'$) is given by
\begin{eqnarray} 
\label{FactorizeAmp}
  \M^{\lambda\lambda'}_{hh'} &=& \sum_{d,i} 
  \int_0^1 dx_1\,dx_2\,dx_3\,dy_1\,dy_2\,dy_3\, 
       \delta\Bigl(1 - \sum_{j=1}^3 x_j \Bigr) 
       \delta\Bigl(1 - \sum_{k=1}^3 y_k \Bigr) \\ \nonumber 
&& \hskip 2 cm \times
  \, \phi_i(\vec{x}) \, T_i^{(d)}(\vec{x},h,\lambda;\vec{y},h',\lambda')
  \, \phi_i^*(\vec{y}) \,,
\end{eqnarray}
where the vectors $\vec{x} \equiv (x_1,x_2,x_3)$ and
$\vec{y} \equiv (y_1,y_2,y_3)$ represent the quark longitudinal momentum
fractions; $i$ labels the independent three-valence-quark Fock states of the 
proton, with distribution amplitudes $\phi_i(\vec{x})$;
and $d$ represents the sum over the diagrams that contribute to 
the hard-scattering amplitude $T_i$.  

The distribution amplitude represents the three-valence-quark component of
the proton's light-cone wave function, after the latter is integrated over
transverse momenta up to a factorization scale $\mu$.  (Moments of the
distribution amplitude can also be defined via the matrix elements of
appropriate local three-quark operators.)  The distribution amplitude
evolves logarithmically with $\mu$, but (as was also done in
refs.~\cite{MF,FZ,KN,GV}) we shall neglect this evolution here.  The full
distribution amplitude for a positive-helicity proton is, in the notation
of ref.~\cite{KN},
\begin{equation}
 \vert p_\uparrow \rangle = {f_N \over 8\sqrt{6}}
 \int_0^1 dx_1\, dx_2\, dx_3\ \delta\Bigl(1 - \sum_{j=1}^3 x_j \Bigr)
  \ \sum_{i=1}^3 \phi_i(\vec{x}) \ \vert i; \vec{x} \rangle \,,
\label{KNstatedef}
\end{equation}
where
\begin{eqnarray}
\nonumber
 \vert 1; \vec{x} \rangle &=& 
\vert u_\uparrow(x_1) u_\downarrow(x_2) d_\uparrow(x_3) \rangle \,, \\  
 \vert 2; \vec{x} \rangle &=& 
\vert u_\uparrow(x_1) d_\downarrow(x_2) u_\uparrow(x_3) \rangle \,, \\  
\nonumber
 \vert 3; \vec{x} \rangle &=& 
\vert d_\uparrow(x_1) u_\downarrow(x_2) u_\uparrow(x_3) \rangle \,.
\end{eqnarray}
The normalization constant $f_N$ can be determined from QCD sum rules or
lattice QCD.  We choose $f_N = 5.2 \times 10 ^{-3}$~GeV$^2$ (as in
refs.~\cite{KN,GV}).  Fermi-Dirac statistics, isospin and spin symmetry result 
in only one independent distribution amplitude, $\phi_1$; the other two 
are given by 
\begin{eqnarray}
  \phi_2(x_1,x_2,x_3) &=& - \phi_1(x_1,x_2,x_3) - \phi_1(x_3,x_2,x_1) \,, \\
 \nonumber \phi_3(x_1,x_2,x_3) &=&   \phi_1(x_3,x_2,x_1)\,. 
\end{eqnarray}
In addition to neglecting evolution of the distribution amplitude, we
shall also take $\alpha_s$ to be fixed.  The Born-level 
cross section then scales as $\alpha_s^4 \times f_N^4$.

The hard scattering amplitude is computed for three collinear incoming 
and outgoing quarks.  The color and electric charge dependence can be 
factored off of each diagram as
\begin{equation}
 T_i^{(d)}(\vec{x},h,\lambda;\vec{y},h',\lambda')
 = C^{(d)} \, g^4 \, Z_i^{(d)}  \, 
   \tilde{T}^{(d)}(\vec{x},\vec{y};h,\lambda,h',\lambda') \,,
\end{equation}
where $C^{(d)}$ is the color factor, $g$ is the strong coupling constant,
and $Z_i^{(d)}$ is the appropriate product of quark electric charges,
while $\tilde{T}^{(d)}$ is color and flavor independent.

The helicities of the quarks in the hard scattering amplitude are
conserved by the gauge interactions; therefore the proton helicity is
conserved, and $\M^{\lambda\lambda'}_{hh'} = 0$ for $h \neq h'$.
Parity and time-reversal invariance further reduce the number of 
independent helicity amplitudes to three, which we take to be
\begin{equation}
  \M^{\uparrow\uparrow}_{\uparrow\uparrow} \,, \quad
  \M^{\uparrow\downarrow}_{\uparrow\uparrow} \,, \quad {\rm and} \quad
  \M^{\downarrow\downarrow}_{\uparrow\uparrow} \,.
\label{ThreeHelAmps}
\end{equation}
In principle, 378 diagrams contribute to the hard scattering amplitude.
However, 42 of them contain three-gluon vertices and have a vanishing
color factor.  Many others vanish for individual helicity configurations.  

We adopted the technique in ref.~\cite{KN} of using the parity symmetry
(denoted ${\cal E}$ therein) between certain classes of diagrams to reduce
the number that had to be computed, while reserving the time-reversal
symmetry as a check.  All diagrams were computed by two independent
computer programs, both based on the formalism outlined in ref.~\cite{FN}.
These expressions were found to be identical to those used in the two most
recent computations~\cite{KN,GV}.\footnote{%
We compared our results for each diagram to the formulae given in 
Tables III and IV of ref.~\cite{KN}. These tables contain three errors 
(found by M.~Vanderhaeghen~\cite{VandPrivate} as well as us) in addition 
to one in diagram $A71$ that was published in an erratum.  However, all 
these errors are typographical and do not affect the numerical results in 
that paper~\cite{KronPrivate}.  The errors are: In the denominator of
$\tilde{T}^{(A44)}(x,\uparrow,\downarrow;y,\uparrow,\downarrow)$,
$(\bar{x}_3,x_1)$ should be $(\bar{x}_3,y_1)$;
$\tilde{T}^{(C75)}(x,\uparrow,\downarrow;y,\uparrow,\downarrow)$ should be
multiplied by $1/c$; and the diagram related to $\overline{C77}$ by ${\cal
T \circ E}$ should be $\overline{F11}$, not $\overline{F33}$.}%
Thus we agree completely with refs.~\cite{KN,GV} on the hard scattering 
amplitude $T_i$.

The next step is to perform the four-dimensional integration in 
\eqn{FactorizeAmp} over the independent quark momentum fractions.
For the various model distribution amplitudes~\cite{CZ,GS,KS,COZ}
we used the coefficients of $\phi_1$ listed in Table I of ref.~\cite{KN}
(and eq. (6) of ref.~\cite{SB}).
Many diagrams include denominators that vanish inside the
$(\vec{x},\vec{y})$ integration region, due to the
presence of an internal quark and/or gluon that can go on shell.  
This is not a true long-distance singularity, and all the integrals are
finite, diagram by diagram, but it is a technical obstacle to obtaining
a reliable value for the integral.  In the notation of ref.~\cite{KN}, 
the Feynman $i\ve$ prescription leads to singular denominators of the form
\begin{equation}
{1 \over (x,y) + i \ve} = {\rm P} { 1 \over (x,y) } 
   - i \pi \delta\bigl( (x,y) \bigr) \,, 
\label{SingDenom}
\end{equation}
where P stands for principal part and $(x,y) \equiv x (1-y s^2) - y c^2$, 
with $s=\sin(\theta/2)$ and $c=\cos(\theta/2)$.  
Diagrams can be classified by the number of singular factors found in 
the denominator; for the Compton process this number can be 0, 1, 2 or 3.
The presence of on-shell partons in the Born-level hard scattering
amplitude (for particular values of $(\vec{x},\vec{y})$) leads to large
phases in the PQCD amplitude~\cite{FZ,MF,KN}.  This is in contrast to the 
handbag model, which predicts an imaginary part that is small and beyond
the accuracy of the model.

At least four different numerical methods have previously been applied to
handle the singular integrations.  Ref.~\cite{MF} performed a Taylor
expansion of the numerators of the integrand symmetrically about each 
singularity.  Ref.~\cite{FZ} kept the $\ve$ in \eqn{SingDenom} explicit, 
and evaluated the integrals for a sequence of $\ve$ values tending to
zero, looking for stable results.  Ref.~\cite{KN} handled the 
imaginary parts of the singular integrals by solving the $\delta$-function
constraint explicitly, and carried out the real, principal-part integrals by 
folding the region of integration over at the singularity, so that the
integrand is manifestly finite.  Finally, ref.~\cite{GV} deformed the
$(\vec{x},\vec{y})$ integration contour into the complex plane, an elegant
technique that requires relatively little bookkeeping for its implementation.

We adopted a variation of the contour deformation technique~\cite{GV}.
We first let
\begin{eqnarray}
&& x_1 = \xi_1, \qquad x_2 = (1-\xi_1)(1-\xi_2), \qquad x_3 = (1-\xi_1)\xi_2, 
\label{ChangeofVar} \\
\nonumber
&& y_1 = \eta_1, \qquad y_2 = (1-\eta_1)(1-\eta_2), 
\qquad y_3 = (1-\eta_1)\eta_2,
\end{eqnarray}
so that the four independent variables $(\xi_1,\xi_2,\eta_1,\eta_2)$ were
integrated on the interval [0,1].  We then deformed the single variable
$\eta_1$ into the complex plane, so that it ran either over the piecewise
linear contour $0 \to i\eps \to 1+i\eps \to 1$, or over a semi-circular
contour extending from 0 to 1.  Note that this simultaneously deforms both
$y_1$ and $y_3$, towards opposite sides of the real axis, while $x_1$ and
$x_3$ remain real.  Inspection of the denominator factors in Tables III
and IV of ref.~\cite{KN} shows that this deformation is sufficient to
correctly bypass the singularities in every Compton diagram.  For example,
the denominator of diagram A16 includes the factors 
$[ (x_1,y_1) + i\ve ] [ (\bar{x}_3,y_1) + i\ve ] [ (y_3,x_3) + i\ve ]$, 
where $\bar{x}_i \equiv 1-x_i$, $\bar{y}_i \equiv 1-y_i$.  
Using the identity $(x,y) = (\bar{y},\bar{x})$, the singular factors can 
be rewritten as 
$[ (1-y_1,\bar{x}_1) + i\ve ] [ (1-y_1,x_3) + i\ve ] 
[ (y_3,x_3) + i\ve ]$, which shows that $y_1$ and $y_3$ should indeed 
be deformed in opposite directions.  
If the diagram happens to contain denominators of the form
$(x_i,y_1)$ or $(y_3,x_i)$, instead of $(y_1,x_i)$ or $(x_i,y_3)$, as does
diagram A16, then the imaginary part should be multiplied by an
overall minus sign (or equivalently, the contour should be deformed in the 
opposite direction with respect to the real axis). 

After making these contour deformations, the real and imaginary parts of 
the complex integrals were performed separately using the Monte Carlo
integration routine VEGAS~\cite{VEGAS}.  Two independent versions of the 
contour integration were implemented numerically, with two different
choices of contour (piecewise linear vs. semi-circular), and we also
varied the deformation parameter $\eps$, obtaining stable results.

A third version of the integration program was constructed, which employed
the Gauss-Legendre formalism~\cite{NumRecipes} with ten points per
integration variable, instead of VEGAS.  Although the Gauss-Legendre
errors were larger than the VEGAS errors, the two sets of results were
completely consistent with each other (and were both inconsistent with
results from previous work; see section~\ref{ComparisonSubSection}).

We carried out further checks on our integration routines.  For diagrams
with only one singular factor in the denominator, one can integrate the
imaginary part analytically.  Using this procedure we checked the
imaginary part of all diagrams with one singularity.  One can also check
the diagrams with no singularities in the same manner.  A second check
employed the identity
\begin{equation}
  \frac{1}{(x,y)(x,z)} = \frac{1-ys^2}{c^2(y-z)(x,y)}
                       - \frac{1-zs^2}{c^2(y-z)(x,z)} \,.
\label{pf}
\end{equation}
Using \eqn{pf} one can reduce all three-singularity diagrams to two
singularities. (These expressions can be reduced no further, though, 
because the four remaining singular variables are all different.)
One can also reduce all diagrams that initially had two singularities
to one-singularity diagrams, allowing their imaginary parts to be computed 
analytically.  Our integration techniques were robust against all of these
tests.
  
Finally, Table V of ref.~\cite{KN} gives detailed results for diagram
$A51$, which has two denominator singularities.  We agree completely with
these results, for both the real and imaginary parts.  We note that
ref.~\cite{KN} also attempted to evaluate this diagram by implementing the
explicit $\ve \rightarrow 0$ method of ref.~\cite{FZ}, but they obtained
very different results for the imaginary part, compared with the results
of their folding method.  Ref.~\cite{KN} claims that the explicit $\ve
\rightarrow 0$ method is not numerically stable.  Since we agree with
their results for diagram A51, we do not have cause to disagree with this
claim.


\section{Results}
\label{ResultSection}

\subsection{Comparison with previous work}
\label{ComparisonSubSection}

We computed the Compton helicity amplitudes for a variety of distribution
amplitudes, which we refer to as CZ \cite{CZ}, GS \cite{GS}, KS \cite{KS},
COZ \cite{COZ}, HET \cite{SB}, and ASY (the distribution amplitude for
asymptotically large energy scales, $\phi_1(x_1,x_2,x_3) =
120x_{1}x_{2}x_{3}$).  The CZ, KS and COZ distribution amplitudes, which
satisfy the constraints imposed by QCD sum rules~\cite{CZ,KS}, are
qualitatively similar.  They feature a peak in $\phi_1$ for $x_1 \approx
1$, $x_{2,3} \approx 0$; that is, the $u$ quark with the same helicity as
the proton carries most of the momentum.  The GS distribution amplitude
has a peak in $\phi_1$ for $x_{1,3} \approx 1/2$, $x_2 \approx 0$; thus it
splits the momentum more equitably between the two quarks carrying the
proton's helicity.  The HET distribution amplitude is intermediate in
shape between GS and the $\{$CZ,KS,COZ$\}$ class.

Before discussing our full results, we present a comparison of results in
the literature.  Here we choose the COZ distribution amplitude, since it
was employed in four of the five existing calculations.  (Only the
earliest calculation~\cite{MF}, which was later superseded~\cite{FZ}, did
not use the COZ distribution amplitude.)  The overall cross-section
normalizations in the literature are sometimes difficult to determine, due
for example to unspecified choices for $\alpha_s$.  Therefore we choose to
compare results for the following (normalization-independent)
initial-state helicity correlation~\cite{GV},
\begin{equation}
 A_{LL}\ \equiv\ { 
{\ds d\sigma_+^+ \over \ds dt} - {\ds d\sigma_+^- \over \ds dt} \over 
{\ds d\sigma_+^+ \over \ds dt} + {\ds d\sigma_+^- \over \ds dt} } 
\ ,
\label{ALL}
\end{equation}
where $d\sigma_h^\lambda/dt$ is the differential cross section for a 
helicity $h$ proton scattering off a helicity $\lambda$ photon.  

\begin{figure}
\begin{center}
\includegraphics[scale=0.75]{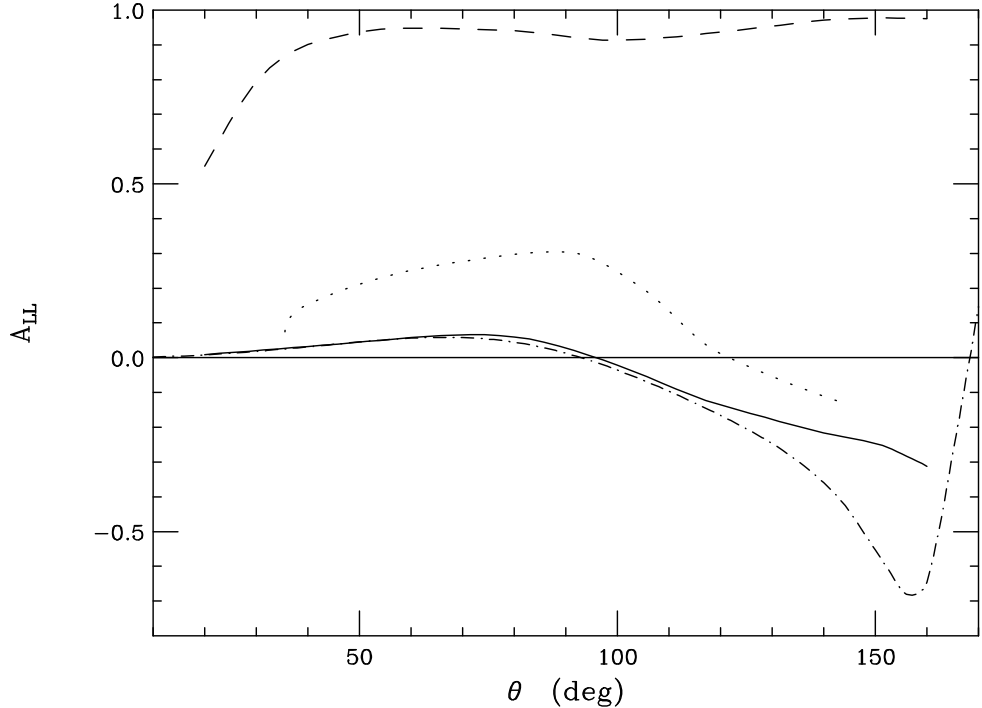}
\caption{\small Four different calculations of the polarization asymmetry
$A_{LL}$ defined in \eqn{ALL}, for the COZ distribution amplitude.  
The dotted line is from ref.~\cite{FZ}, the dashed line from
ref.~\cite{KN}, the dot-dash line from ref.~\cite{GV}, and the solid line 
from this work.
\label{ALLallFigure}}
\end{center}
\end{figure}
 
\Fig{ALLallFigure} shows that none of the four calculations of $A_{LL}$ 
agrees completely with any other.  The only two results that are very
close are ours and that of ref.~\cite{GV}.  These two curves are in excellent 
agreement for $\theta < 110^\circ$; however, we do not reproduce the 
prominent dip of ref.~\cite{GV} in the backward region.  This statement
is true for the four distribution amplitudes we have compared: KS, COZ,
CZ and ASY~\cite{GV,VandPrivate}.  Figure~14 of the second
reference in~\cite{GV} shows that the dip in $A_{LL}$ derives from 
${d\sigma\over dt}(\gamma_\uparrow p_\uparrow \to \gamma p)
\propto | \M^{\uparrow\uparrow}_{\uparrow\uparrow} |^2 + 
        | \M^{\uparrow\downarrow}_{\uparrow\uparrow} |^2$,
and not from 
${d\sigma\over dt}(\gamma_\downarrow p_\uparrow \to \gamma p)
\propto | \M^{\uparrow\downarrow}_{\uparrow\uparrow} |^2 + 
        | \M^{\downarrow\downarrow}_{\uparrow\uparrow} |^2$.
Indeed, we agree with their 
$\gamma_\downarrow p_\uparrow \to \gamma p$
cross section for all angles to better than 10\%, up to an overall 
normalization factor which can be accounted for by different choices for 
$\alpha_s$.  We agree with the $\gamma_\uparrow p_\uparrow \to \gamma p$
cross section only for $\theta < 110^\circ$, however.  This suggests that 
the discrepancy with ref.~\cite{GV} is predominantly from the single 
helicity amplitude $\M^{\uparrow\uparrow}_{\uparrow\uparrow}$.  

The curve from ref.~\cite{FZ} has the same general shape as ours, but is 
offset from it.  The phases of the dominant helicity amplitudes given 
in ref.~\cite{FZ} actually agree quite well with our results in 
figs.~\ref{UUPhaseFigure}--\ref{DDPhaseFigure} below; the magnitudes
are offset by relatively angle-independent factors.

Ref.~\cite{KN} finds a very large asymmetry.  We have made a detailed
comparison of our COZ results with those of ref.~\cite{KN}, for the real
and imaginary parts of the three independent helicity amplitudes.  Each
amplitude has been further split into four pieces~\cite{KronPrivate},
according to the number of singular propagators in the diagram (as
determined from Tables III and IV of ref.~\cite{KN}).  The zero propagator
terms (which were integrated analytically by both groups) agree to high
precision (6 digits).  The one propagator terms agree to within VEGAS
errors, except for the imaginary part of one helicity amplitude
($\M^{\uparrow\downarrow}_{\uparrow\uparrow}$) which is within 10\%.  For
the two propagator terms, we are in agreement on the real part of
$\M^{\uparrow\uparrow}_{\uparrow\uparrow}$ and
$\M^{\uparrow\downarrow}_{\uparrow\uparrow}$, but have a large discrepancy
in the imaginary part.  Strangely enough, for
$\M^{\downarrow\downarrow}_{\uparrow\uparrow}$ we agree on the imaginary
part but disagree on the real part!  For the three propagator terms, both
the real and imaginary parts disagree for all three helicity amplitudes.
The bulk of our overall numerical disagreement comes from the two
propagator terms contributing to the imaginary part of
$\M^{\uparrow\uparrow}_{\uparrow\uparrow}$.  The two propagator terms are
often 100 times larger than ours, and they drive ref.~\cite{KN}'s values 
for ${\rm Im}\, \M^{\uparrow\uparrow}_{\uparrow\uparrow}$ to be roughly a
factor of 10 larger than ours.

We also calculated $A_{LL}$ for the COZ distribution amplitude using
Gauss-Legendre integration instead of VEGAS.  The result agrees with our
VEGAS result shown in \fig{ALLallFigure} (albeit with larger errors), and
it disagrees with the other results, in particular that of ref.~\cite{GV} 
for $\theta > 110^\circ$.


\subsection{Helicity amplitudes and unpolarized cross section}
\label{HelCrossSubSection}


\begin{figure}
\begin{center}
\includegraphics[scale=0.75]{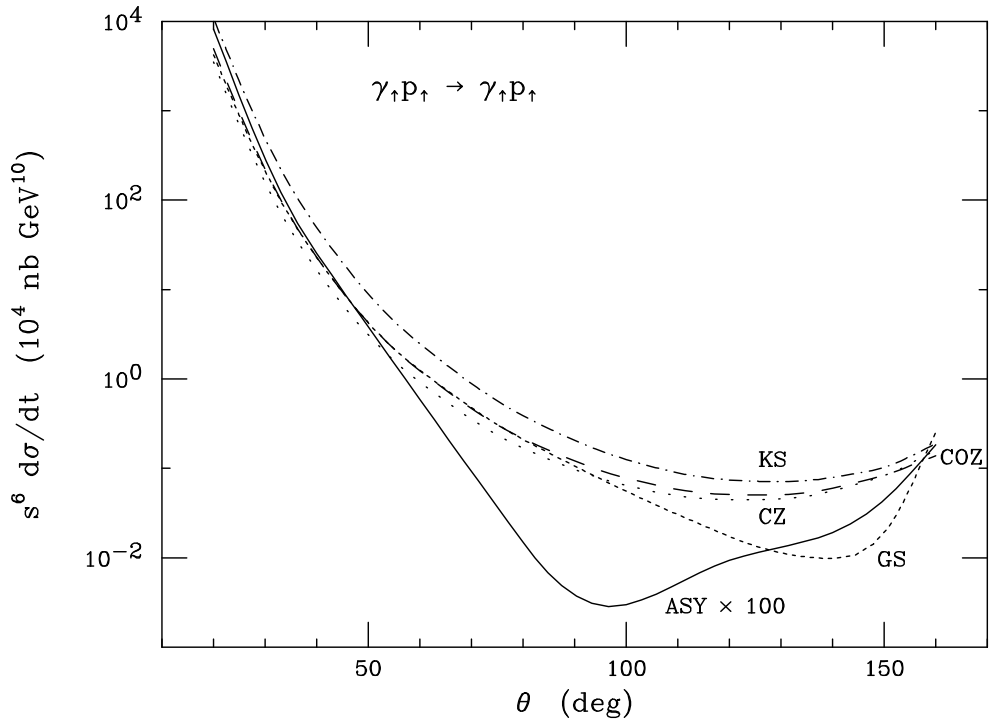}
\caption{\small The cross section for 
$\gamma_\uparrow p_\uparrow \rightarrow \gamma_\uparrow p_\uparrow$
for five different distribution amplitudes, CZ, COZ, KS, GS and ASY.
The results for the asymptotic distribution amplitude (ASY) have been 
multiplied by 100.
\label{UUCrossFigure}}
\end{center}
\end{figure}

\begin{figure}
\begin{center}
\includegraphics[scale=0.75]{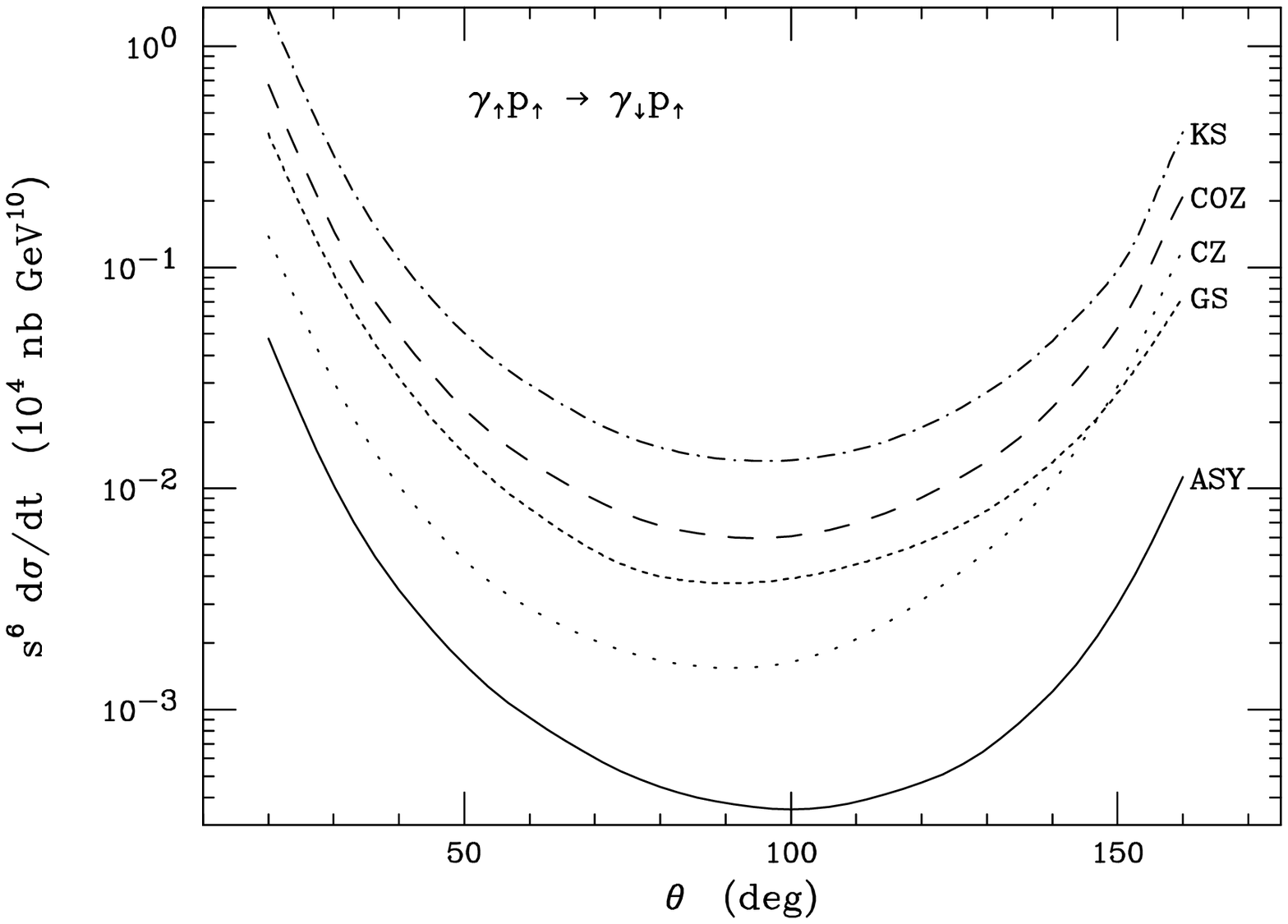}
\caption{\small The cross section for 
$\gamma_\uparrow p_\uparrow \rightarrow \gamma_\downarrow p_\uparrow$.
\label{UDCrossFigure}}
\end{center}
\end{figure}

\begin{figure}
\begin{center}
\includegraphics[scale=0.75]{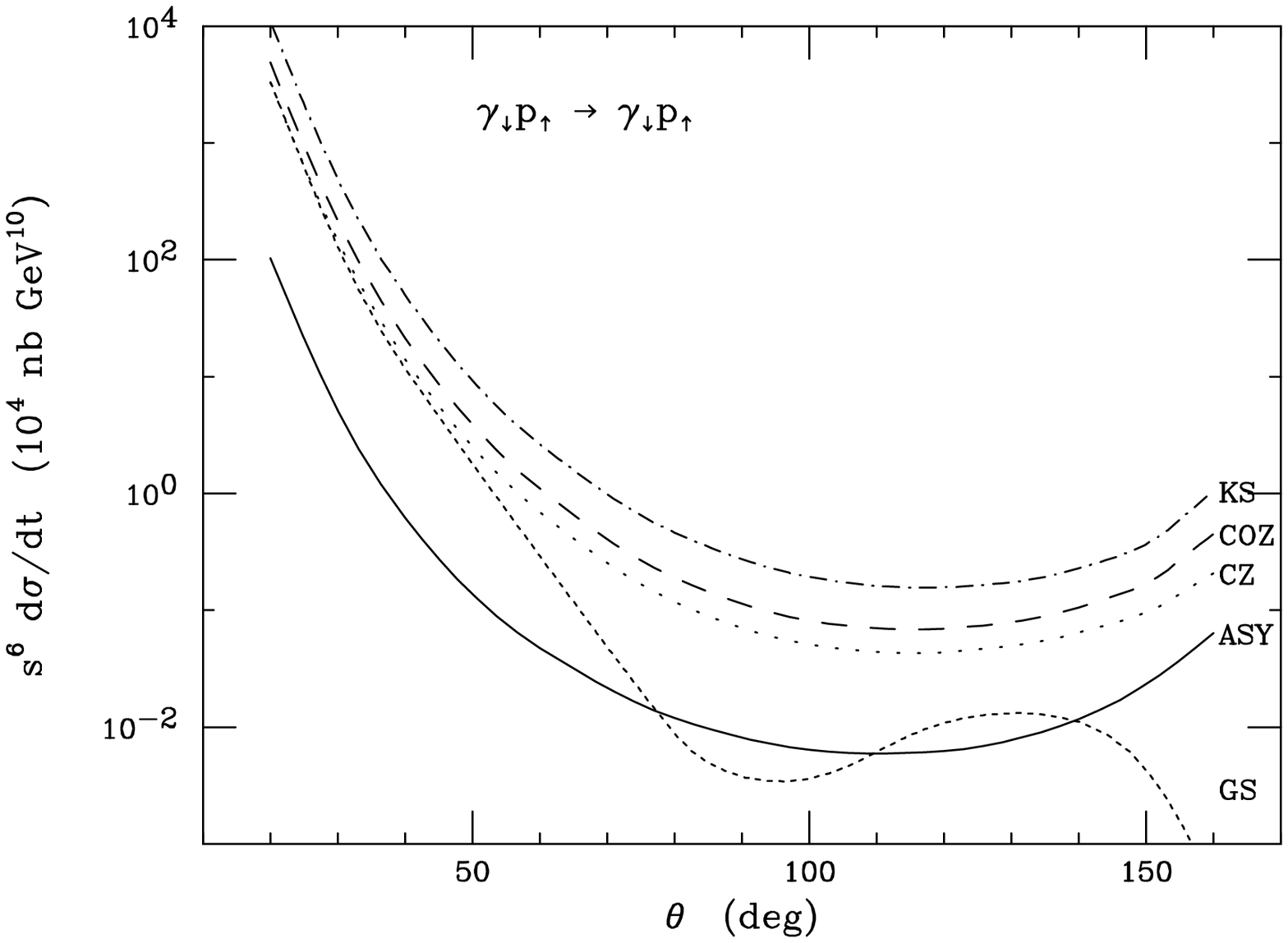}
\caption{\small The cross section for 
$\gamma_\downarrow p_\uparrow \rightarrow \gamma_\downarrow p_\uparrow$.
\label{DDCrossFigure}}
\end{center}
\end{figure}

\begin{figure}
\begin{center}
\includegraphics[scale=0.75]{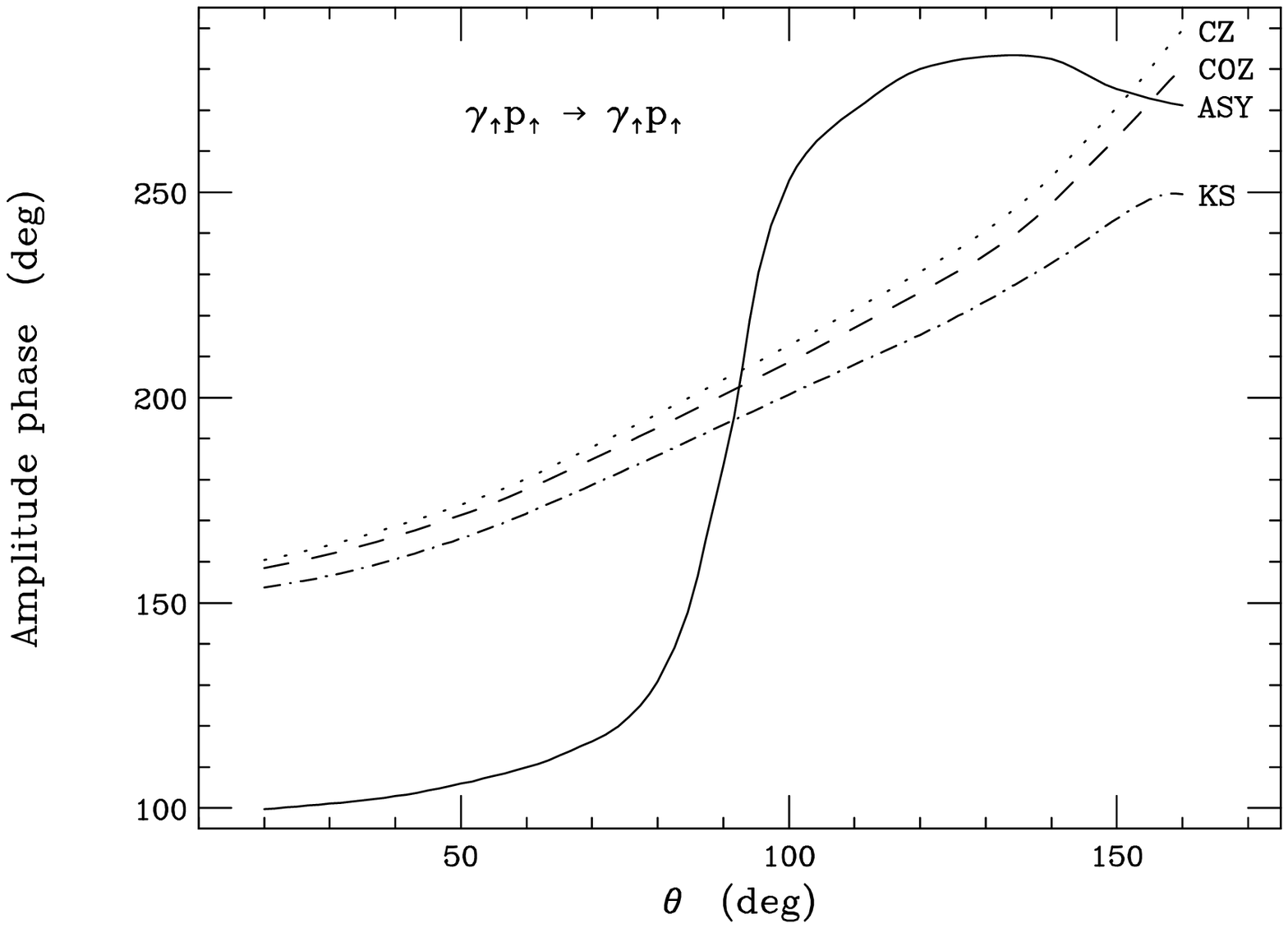}
\caption{\small Phase of the helicity amplitude for 
$\gamma_\uparrow p_\uparrow \rightarrow \gamma_\uparrow p_\uparrow$
for the distribution amplitudes CZ, COZ, KS and ASY.
\label{UUPhaseFigure}}
\end{center}
\end{figure}

\begin{figure}
\begin{center}
\includegraphics[scale=0.75]{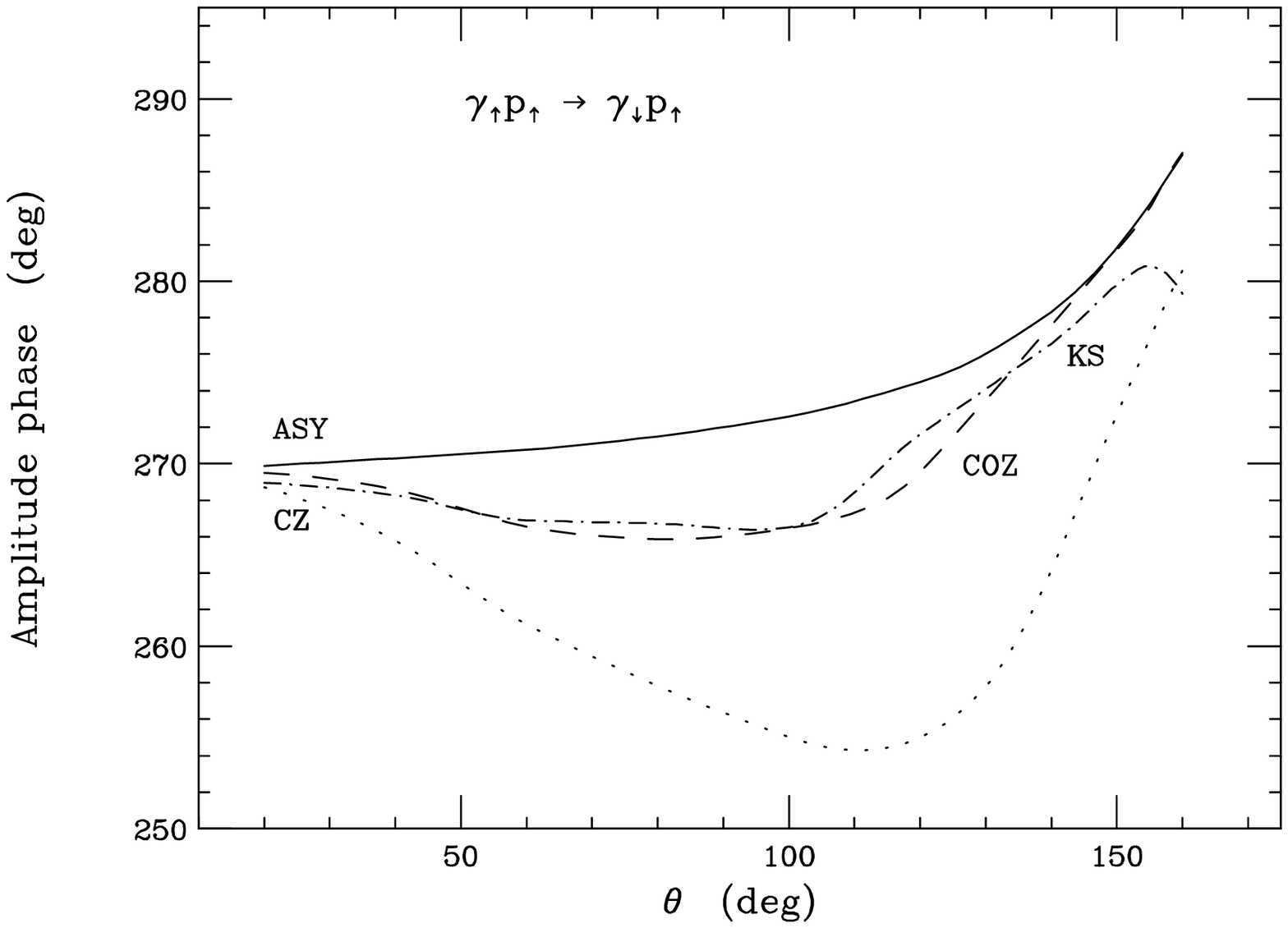}
\caption{\small Phase of the helicity amplitude for 
$\gamma_\uparrow p_\uparrow \rightarrow \gamma_\downarrow p_\uparrow$.
\label{UDPhaseFigure}}
\end{center}
\end{figure}

\begin{figure}
\begin{center}
\includegraphics[scale=0.75]{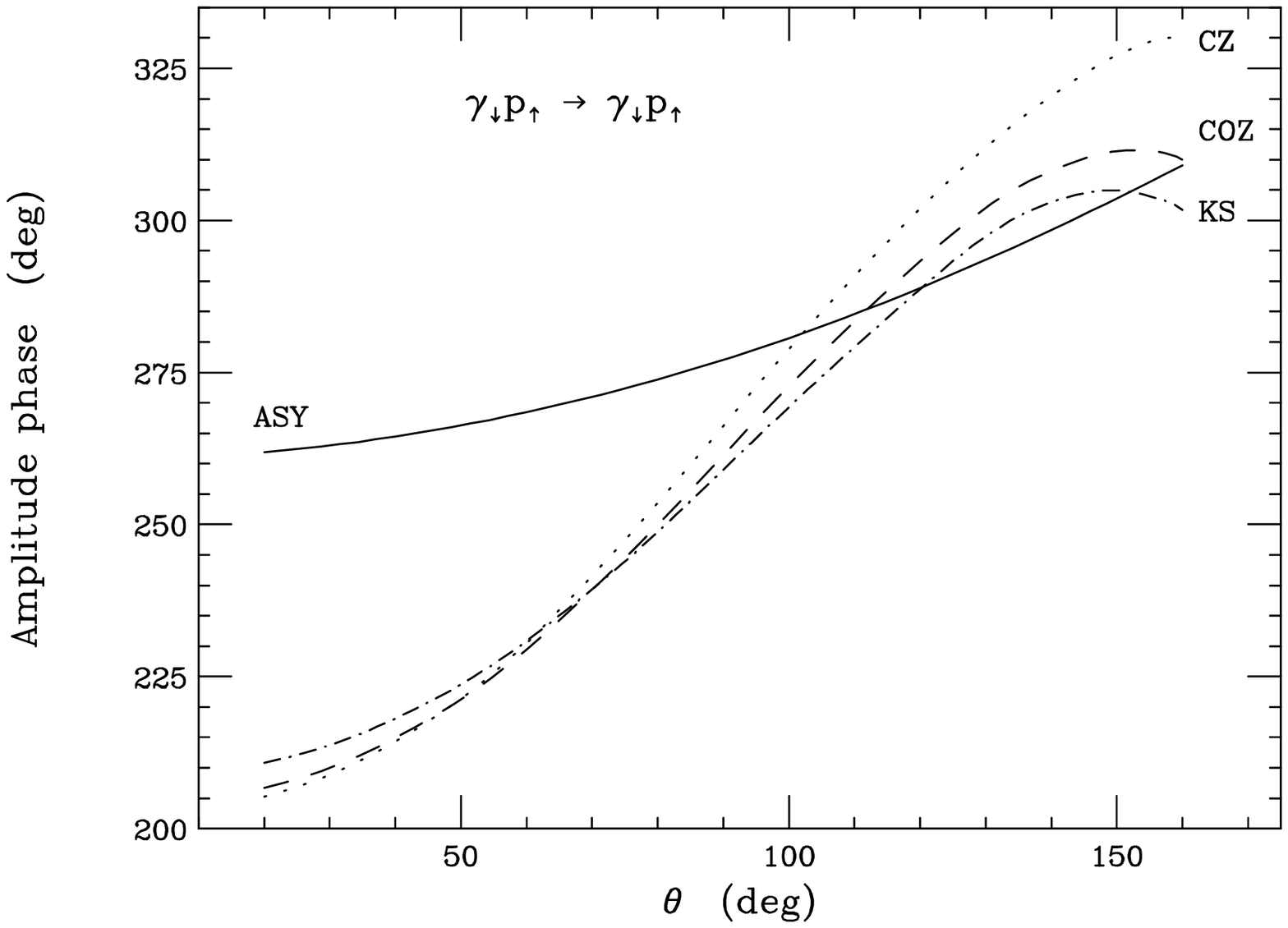}
\caption{\small Phase of the helicity amplitude for 
$\gamma_\downarrow p_\uparrow \rightarrow \gamma_\downarrow p_\uparrow$.
\label{DDPhaseFigure}}
\end{center}
\end{figure}

\begin{figure}
\begin{center}
\includegraphics[scale=0.75]{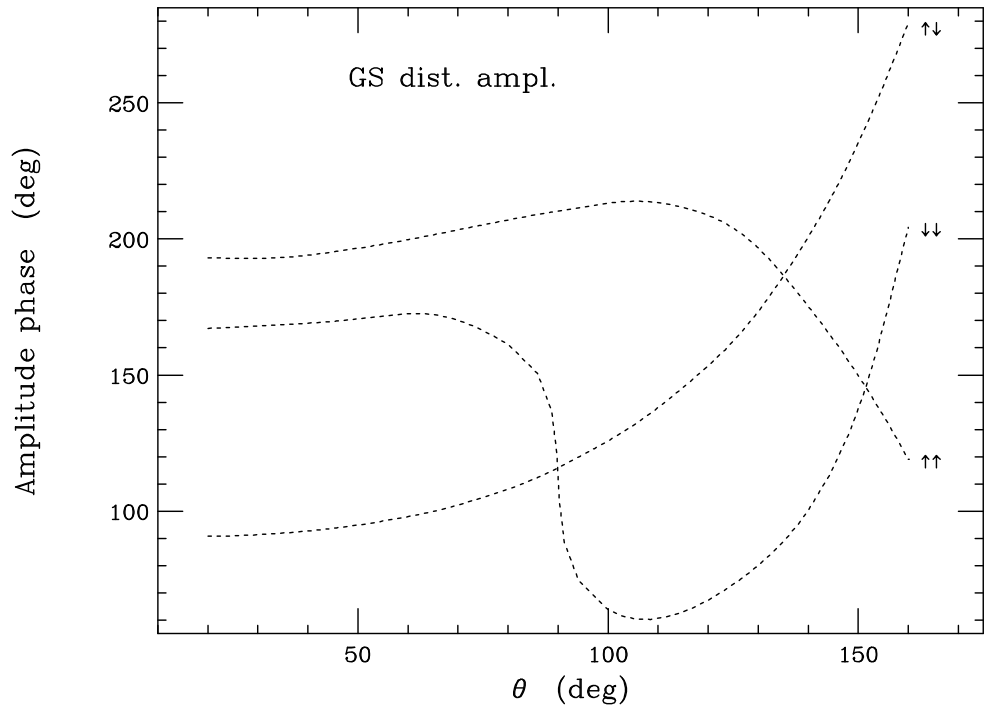}
\caption{\small Phase of the three independent helicity amplitudes for
the GS distribution amplitude.  The arrows correspond to the photon
helicities $\lambda$, $\lambda'$ in the amplitudes
$\M^{\lambda\lambda'}_{\uparrow\uparrow}$.
\label{GSPhaseFigure}}
\end{center}
\end{figure}

In figs.~\ref{UUCrossFigure}--\ref{DDCrossFigure}
we display our results for the polarized differential 
cross sections, 
\begin{equation} 
s^6 \, { d \sigma_{hh'}^{\lambda\lambda'} \over dt} 
= { s^4 \over 16 \pi } \, | \M_{hh'}^{\lambda\lambda'} |^2 \,,
\label{polxs}
\end{equation}
for the three independent helicity configurations.  Each figure
plots the results for five different distribution amplitudes.
(For HET we shall only plot the unpolarized cross section.)
These plots were made for $\alpha^{-1}_{\rm em} = 137.036$, 
$\alpha_s = 0.3$ and $f_N = 5.2 \times 10^{-3}$~GeV$^2$, so they can 
be compared directly with ref.~\cite{KN}.  The phases of the helicity 
amplitudes are plotted in 
figs.~\ref{UUPhaseFigure}--\ref{DDPhaseFigure}; the GS distribution
amplitude has a much different behavior and is therefore plotted
separately, in \fig{GSPhaseFigure}.  The phases are generally large;
indeed $\M^{\uparrow\downarrow}_{\uparrow\uparrow}$ is almost pure
imaginary (except for the GS distribution amplitude).  For reference, we also
provide in Table~1 our numerical results for the real and imaginary part 
of $\cal{M}^{\uparrow\uparrow}_{\uparrow\uparrow}$, for the COZ 
distribution amplitude, including errors from the VEGAS integration.


\begin{table}
\label{COZupupTable}
\begin{center}
\begin{tabular}[c]{|c|c|c|}\hline
$\theta$ (deg) 
& $10^3 \, s^2 \, {\rm Re}(\cal{M}^{\uparrow\uparrow}_{\uparrow\uparrow})$  
& $10^3 \, s^2 \, {\rm Im}(\cal{M}^{\uparrow\uparrow}_{\uparrow\uparrow})$ 
\\ \hline
20 &  $-$74920 $\pm$  240   & 29200  $\pm$ 230 \\
30 &  $-$15720 $\pm$  110   &  5133  $\pm$ 46 \\
40 &   $-$5255 $\pm$   15   &  1301  $\pm$ 14 \\
50 &   $-$2371.2 $\pm$  8.0 &   348.6  $\pm$ 5.8 \\
60 &   $-$1273.6 $\pm$  4.3 &    42.2  $\pm$ 3.5 \\
70 &    $-$768.8 $\pm$  2.3 &   $-$72.1  $\pm$ 2.3 \\
80 &    $-$511.2 $\pm$  3.1 &  $-$115.4 $\pm$  1.4 \\
90 &    $-$369.8 $\pm$  1.2 &  $-$139.0  $\pm$ 1.0 \\
100 &   $-$278.3 $\pm$  1.0 &  $-$152.03 $\pm$ 0.91 \\
110 &   $-$222.4 $\pm$  1.2 &  $-$165.53 $\pm$ 0.90 \\
120 &   $-$179.54 $\pm$ 0.70 &  $-$183.15 $\pm$ 0.95 \\
130 &   $-$144.2 $\pm$  1.0 &  $-$211.6  $\pm$ 1.0 \\
140 &   $-$107.80 $\pm$ 0.91 & $-$257.1  $\pm$ 1.3 \\
150 &    $-$52.9 $\pm$  2.7 &  $-$324.9  $\pm$ 2.6 \\
160 &     75.9 $\pm$  3.1 &  $-$415.5  $\pm$ 5.0 \\ \hline
\end{tabular}
\end{center}
\caption{\small The real and imaginary parts of the helicity amplitude
$\cal{M}^{\uparrow\uparrow}_{\uparrow\uparrow}$ for the COZ distribution
amplitude (multiplied by $s^2$ in units of GeV$^4$).  The errors are from
the VEGAS numerical integration.  The values used for $f_N$, 
$\alpha_{\rm em}$, and $\alpha_s$ are the same as in the rest of the 
paper.  The normalization is the same as in Table V of ref.~\cite{KN} 
(which we found quite useful).}
\end{table}


\begin{figure}
\begin{center}
\includegraphics[scale=0.75]{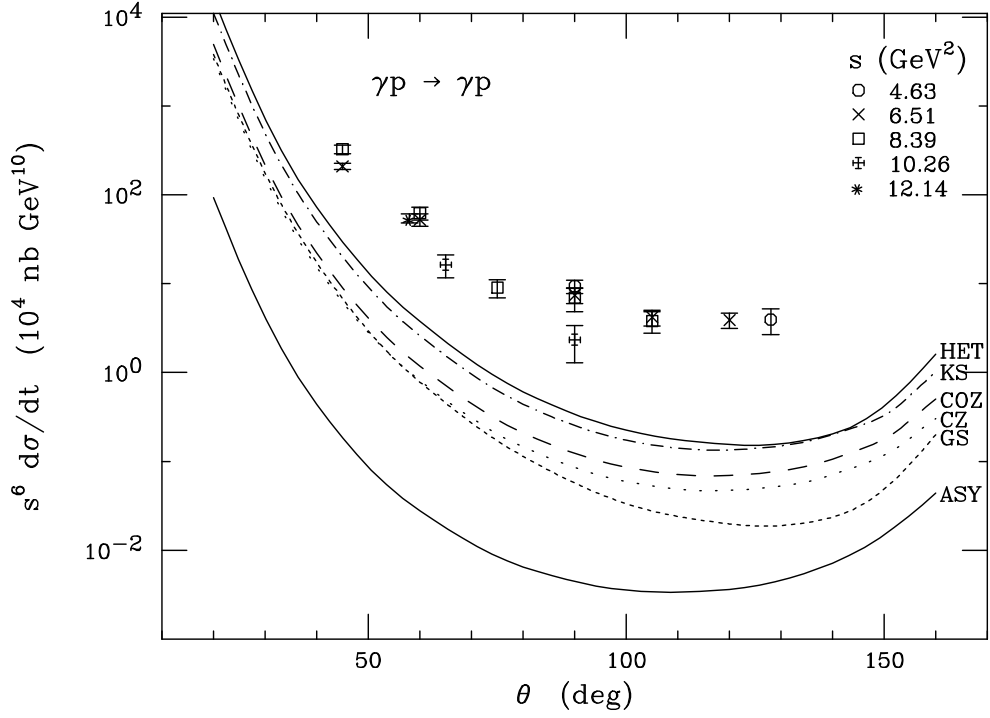}
\caption{\small The unpolarized scaled cross section~(\ref{unpolxs})
for all six distribution amplitudes, for $\alpha_s = 0.3$ and 
$f_N = 5.2 \times 10^{-3}$~GeV$^2$, compared with experiment~\cite{Shupe}.
\label{UnpolFigure}}
\end{center}
\end{figure}

\Fig{UnpolFigure} shows our predictions for the unpolarized differential
Compton cross section, given by
\begin{equation} 
s^6 \frac{d\sigma}{dt} = { 1 \over 4 } \, 
\sum_{\lambda,\lambda',h,h'}
s^6 \, { d \sigma_{hh'}^{\lambda\lambda'} \over dt} \,,
\label{unpolxs}
\end{equation}
along with the experimental data from ref.~\cite{Shupe}.
For the values used $\alpha_s = 0.3$, $f_N = 5.2 \times 10^{-3}$~GeV$^2$,
the predictions lie at least an order of magnitude below the data.
Since the PQCD cross section scales like $\alpha_s^4$,
accommodating a factor of 10 by changing $\alpha_s$ would require 
$\alpha_s \approx 0.5$.  While this is not out of the question, and while
some variation in $f_N$ could be considered as well,
this may be pushing the validity of perturbation theory. 
On the other hand, the {\it shape} of the curves (i.e., ignoring the 
overall normalization) matches the data quite well for the KS, COZ, CZ,
and HET distribution amplitudes.


\subsection{Normalization by $F_1^p(Q^2)$}
\label{NormSubSection}

As mentioned in the introduction, the $\alpha_s^4(\mu)$ scaling of the 
proton Compton cross section at Born level introduces a large 
normalization uncertainty into the PQCD prediction.  Uncertainty in 
$f_N$ also contributes.  Both of these uncertainties can be removed at
Born level by considering the dimensionless ratio~\cite{BrodskyPrivate}
\begin{equation} 
{ s^6 \, {\ds d\sigma_{\gamma p} \over \ds dt} 
  \over [ Q^4 \, F_1^p(Q^2) ]^2 }
\,,
\label{normunpolxs}
\end{equation}
where $F_1^p(Q^2)$ is the elastic Dirac form factor for the proton at
space-like momentum transfer $Q$.  One might also imagine normalizing the
Compton cross section by the time-like proton form factor.  At leading
order in $\alpha_s$, the PQCD predictions in the space-like and time-like
regions are identical~\cite{LB}; however, experimentally the time-like
form factor is larger by a factor of about
two~\cite{TimelikeFF,SpacelikeFF}.  Higher order PQCD corrections can in
principle account for this factor, as Sudakov effects are different in the
two regions~\cite{GoussetPire}.  The Compton scattering kinematics are
much closer to those of the space-like proton form factor than the
time-like one, at least as far as the proton is concerned.  Therefore
Sudakov and related higher-order effects are best cancelled by normalizing
with the space-like form factor.

At leading twist, $F_1^p(Q^2)$ is predicted to be the same as the magnetic
form factor $G_M^p(Q^2)$.  Experimentally, these are close but not
identical~\cite{SpacelikeFF}.  To normalize the experimental Compton
points, we use the experimental form factor values,
\begin{equation} 
 Q^4 F_1^p(Q^2) \approx Q^4 G_M^p(Q^2) \approx 1.0\ {\rm GeV}^4,
\hskip 2cm Q^2 \approx 7\,\hbox{--}\, 15\ {\rm GeV}^2,
\label{exptlFF}
\end{equation}
which are representative of the region where both scaled form factors
flatten out, and are also similar to the highest experimental values of 
$s$ available in Compton scattering.\footnote{%
If one equates the four-momentum transfer to the proton in the two processes 
--- $Q^2$ in the form factor and $-t$ in Compton scattering --- then
the corresponding Compton $s = 2Q^2/(1-\cos\theta)$ should actually be
considerably bigger than $Q^2$. At $90^\circ$, for example, $s=2Q^2$.  
Unfortunately, there are no experimental Compton data with 
$s$ this large (all have $-t < 5.3$~GeV$^2$), so there is not a good
overlap with the region~(\ref{exptlFF}) where the elastic form factor is 
beginning to scale properly.}%
\ To normalize the theoretical Compton curves, we recalculated the proton 
form factor at leading order in PQCD, obtaining
\begin{equation} 
 Q^4 F_1^p(Q^2) = {(4\pi \, \alpha_s \, f_N)^2 \over 216} \, I_F,
\label{OurFF}
\end{equation}
where
\begin{equation} 
 I_F = \cases{ 2.500 \times 10^5  &(CZ), \cr
               2.505 \times 10^5  &(GS), \cr
               3.653 \times 10^5  &(KS), \cr
               2.897 \times 10^5  &(COZ), \cr
               3.303 \times 10^5  &(HET), \cr
               0                  &(ASY). \cr }
\label{OurCn}
\end{equation}
These results, using the wave function~(\ref{KNstatedef}) which is
equivalent to that in ref.~\cite{COZ}, are precisely a factor of
two smaller than several previous calculations using the same wave
functions~\cite{FFPrevious}.  We do not understand the origin of this
discrepancy.  We do agree with the normalization of the hard scattering
amplitude and the form factor in ref.~\cite{CarlsonGross} (which uses,
however, a different representation of the proton wave function than
eq.~(\ref{KNstatedef})).


\begin{figure}
\begin{center}
\includegraphics[scale=0.75]{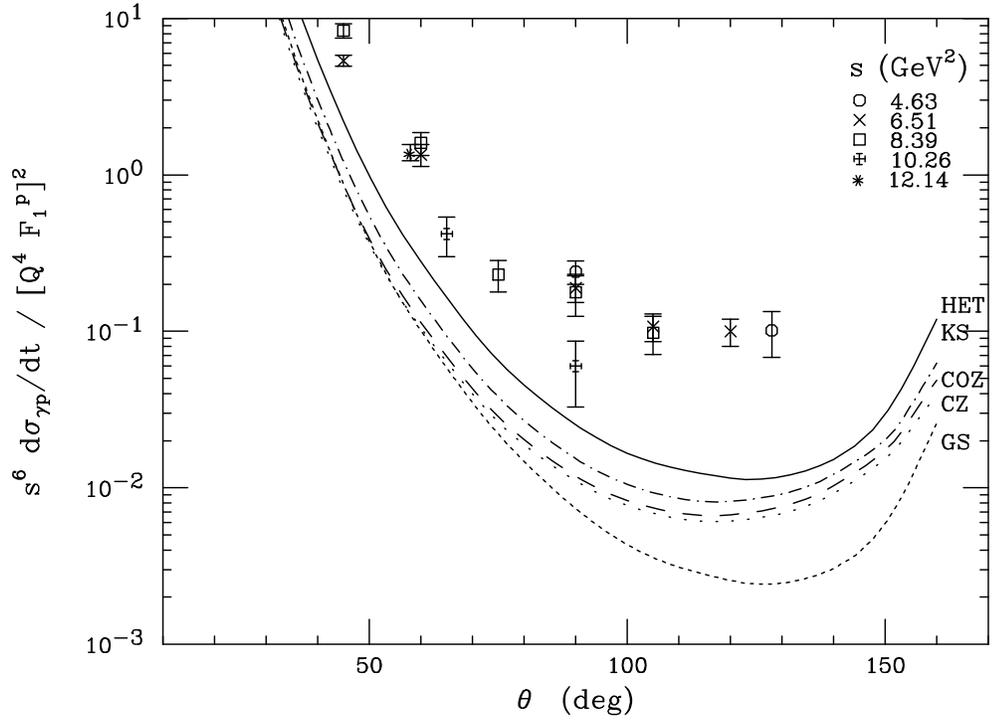}
\caption{\small The scaled unpolarized Compton cross section, normalized
by the scaled elastic proton form factor, as in eq.~(\ref{normunpolxs}),
for five distribution amplitudes, compared with the experimental
data~\cite{Shupe,SpacelikeFF}.
\label{NormUnpolFigure}}
\end{center}
\end{figure}

\Fig{NormUnpolFigure} shows the Compton cross section, normalized
according to eq.~(\ref{normunpolxs}), for both PQCD and the experimental
data.  We omit the ASY distribution amplitude, since the leading order 
ASY form factor vanishes.  Compared with the conventionally normalized
curves in \fig{UnpolFigure}, the spread between the predictions of the three 
qualitatively similar distribution amplitudes, KS, COZ and CZ, has become 
much smaller.  The theoretical curves also lie a factor of 2 to 5 closer
to the data.  However, they still fall about an order of magnitude below
the data at the widest scattering angles.  (The HET distribution amplitude
does slightly better than this.)  Thus it seems unlikely that
the elastic proton form factor and the Compton scattering amplitude
are both described by PQCD at presently accessible energies, unless there
are large higher-order and process-dependent corrections.


\subsection{Asymmetries}
\label{AsymSubSection}

Various polarization asymmetries can be constructed
from the helicity amplitudes.  These observables may provide additional 
diagnostic power for uncovering the Compton scattering mechanism, beyond 
what the unpolarized cross section provides.

\begin{figure}
\begin{center}
\includegraphics[scale=0.75]{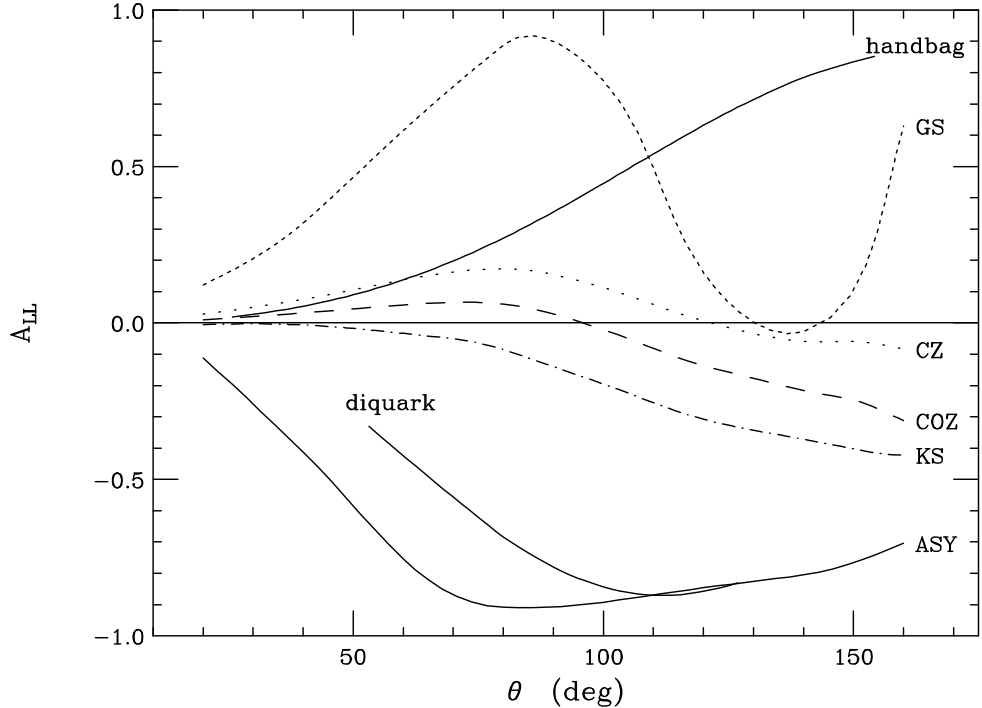}
\caption{\small The initial state helicity correlation $A_{LL}$ in
perturbative QCD for five distribution amplitudes.   Also plotted
is the handbag model prediction for $E_\gamma=4$~GeV (GRV)~\cite{Diehl2},
and a diquark model prediction~\cite{DiQuark}.
\label{ALLFigure}}
\end{center}
\end{figure}

\Fig{ALLFigure} presents the perturbative QCD results for the initial
state helicity correlation $A_{LL}$ defined in \eqn{ALL}.  Also shown is
the handbag model prediction~\cite{Diehl2} for $E_\gamma=4$~GeV, where the
form factors $R_{V,A}$ were evaluated using the parton distribution
functions of GRV~\cite{GRV}.  In leading-twist PQCD, the proton helicity
is conserved.  The handbag model does not inherently require proton
helicity conservation, but it has been assumed in ref.~\cite{Diehl2}.
Thus the PQCD and handbag curves for $A_{LL}$ in \fig{ALLFigure} can be
equated to the longitudinal photon-to-proton polarization transfer
asymmetry, which is slated to be measured for at least one scattering
angle in an upcoming experiment~\cite{JLab}.  The diquark model analyzed
in ref.~\cite{DiQuark,BergerThesis} has nonvanishing proton helicity-flip
amplitudes at finite $s$, making $A_{LL}$ and the polarization transfer
into distinct asymmetries.  We plot the diquark prediction for $A_{LL}$
from ref.~\cite{DiQuark}.  \Fig{ALLFigure} shows that PQCD gives quite
different qualitative behavior from both the handbag and diquark models
for $A_{LL}$, and they should be distinguishable with the help of
experimental data at just a couple of backward scattering angles.  A
caveat is that the GS curve is somewhat oscillatory, so one might wonder
whether a distribution amplitude `between' GS and the $\{$CZ,COZ,KS$\}$
class of amplitudes could produce behavior similar to the handbag model.

\begin{figure}
\begin{center}
\includegraphics[scale=0.75]{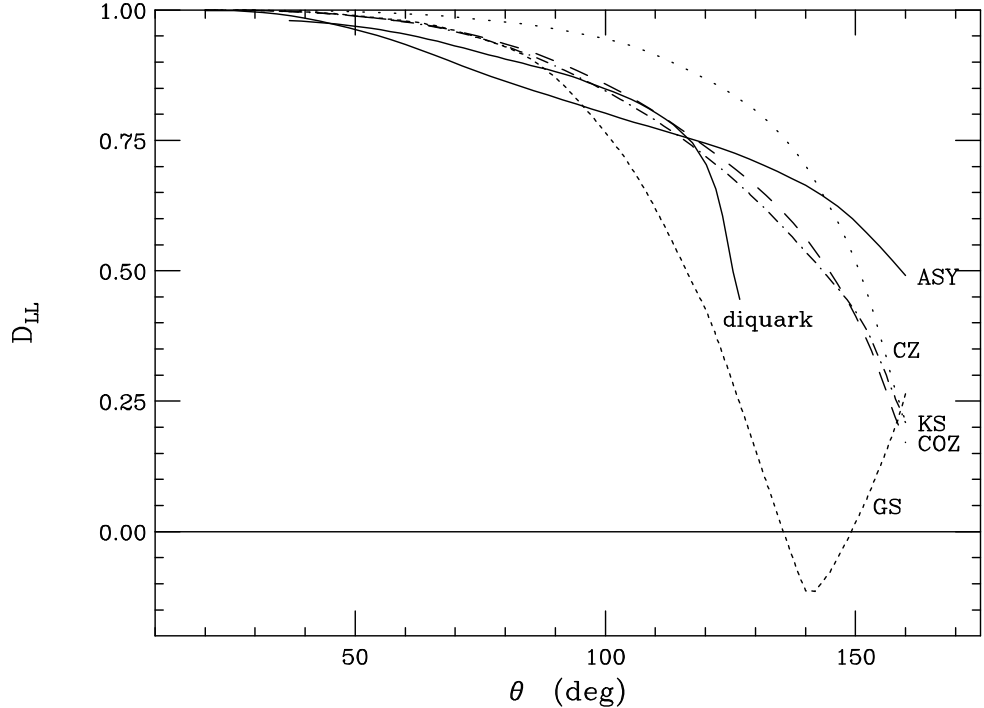}
\caption{\small The photon spin transfer coefficient $D_{LL}$ in
perturbative QCD for five distribution amplitudes.   Also plotted
is the diquark model prediction for $E_\gamma=4$~GeV 
(standard DA)~\cite{BergerThesis}.  The handbag model predicts $D_{LL}=1$.
\label{DLLFigure}}
\end{center}
\end{figure}

One can also define~\cite{BergerThesis} a photon spin transfer coefficient
\begin{equation}
 D_{LL}\ \equiv\ { 
 {\ds d\sigma^{++} \over \ds dt} - {\ds d\sigma^{+-} \over \ds dt}
\over {\ds d\sigma^{++} \over \ds dt} + {\ds d\sigma^{+-} \over \ds dt} } 
\ ,
\label{DLL}
\end{equation}
where now $d\sigma^{\lambda\lambda'}/dt$ is the differential cross
section for initial and final state photon helicities $\lambda$ and 
$\lambda'$, and unpolarized incoming and outgoing protons.
\Fig{DLLFigure} gives the PQCD predictions for this asymmetry, as well
as that of the diquark model for $E_\gamma = 4$~GeV and a `standard'
distribution amplitude~\cite{BergerThesis}.  The handbag model predicts
$D_{LL} = 1$, basically because the helicity-flip quark Compton amplitude 
$\gamma_\uparrow q \to \gamma_\downarrow q$ vanishes at Born level for
massless quarks.  

\begin{figure}
\begin{center}
\includegraphics[scale=0.75]{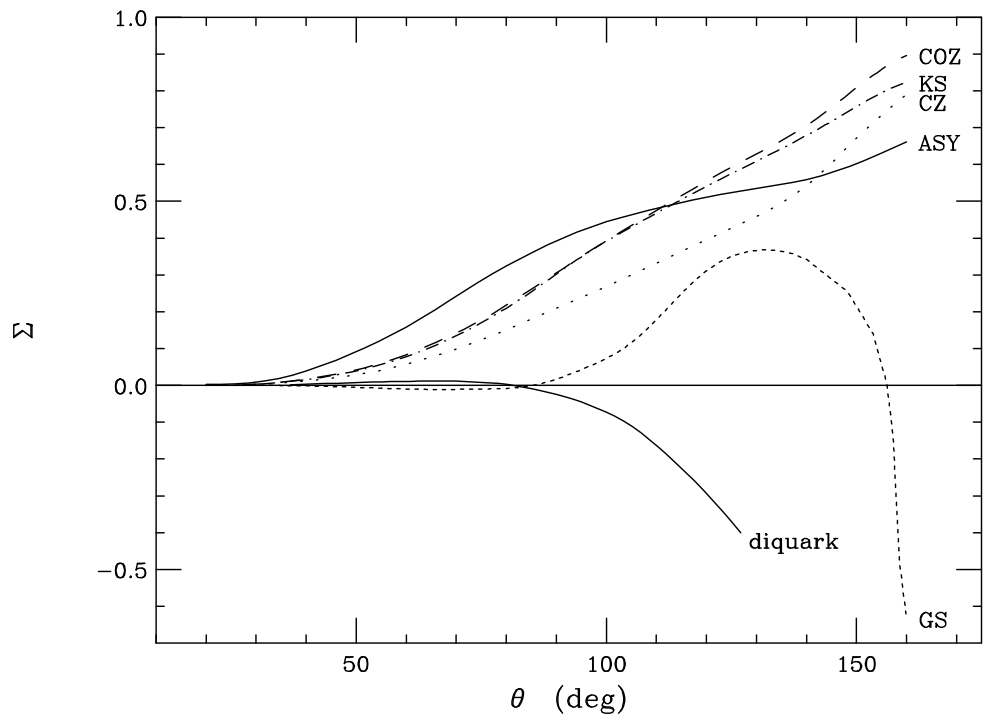}
\caption{\small The photon asymmetry $\Sigma$ in
perturbative QCD for five distribution amplitudes.   Also plotted
is the diquark model prediction for $E_\gamma=4$~GeV 
(standard DA)~\cite{BergerThesis}.  The handbag model predicts $\Sigma=0$.
\label{PhotAsymFigure}}
\end{center}
\end{figure}

The final asymmetry we plot is the photon asymmetry~\cite{BergerThesis} 
\begin{equation}
 \Sigma\ \equiv\ { 
{\ds d\sigma_\perp \over \ds dt} - {\ds d\sigma_\parallel \over \ds dt}
\over {\ds d\sigma_\perp \over \ds dt} 
  + {\ds d\sigma _\parallel\over \ds dt} } \ ,
\label{PhotAsym}
\end{equation}
where $d\sigma_\perp / dt$ and $d\sigma_\parallel / dt$ 
are the differential cross sections for linearly polarized photons, 
with the polarization plane perpendicular or parallel (respectively)
to the scattering plane.  Generation of this asymmetry requires a 
nonzero photon helicity-flip amplitude; hence the asymmetry vanishes in
the handbag model.  \Fig{PhotAsymFigure} plots the PQCD and diquark
predictions.  The diquark prediction is shown for $E_\gamma=4$~GeV 
and a `standard' distribution amplitude; for another distribution 
amplitude $\Sigma$ can become positive in the backward region instead of 
negative~\cite{BergerThesis}.  This asymmetry has actually been
measured~\cite{Buschhorn}, however only for $E_\gamma = 3.45$~GeV and 
$\cos\theta > 0.8$.  A high-energy wide-angle measurement would be very
useful for distinguishing between handbag and PQCD mechanisms.


\section{Conclusions}
\label{ConclusionSection}

Motivated by conflicting results in the literature, we have recalculated
the fixed-order, Born level predictions of perturbative QCD for proton
Compton scattering, for five different distribution amplitudes.  While our
results do not agree with those of any previous group, they do agree very
well with those of ref.~\cite{GV} for $\theta < 110^\circ$, and the
differences for $\theta > 110^\circ$ seem to be dominated by a single
helicity amplitude, $\cal{M}^{\uparrow\uparrow}_{\uparrow\uparrow}$.

From the helicity amplitudes we computed three separate polarization
asymmetries.  Experimental measurements of these asymmetries could be used
in conjunction with the unpolarized differential cross section in order to
help shed light on the mechanism involved in the Compton scattering
process.

We also have attempted to reduce the uncertainty in the overall
normalization of the Compton cross section by normalizing it by the square
of the elastic proton form factor.  This exercise reduces the spread in
the theoretical predictions, but it leaves them an order of magnitude
below the data.  Unfortunately, this result makes it difficult to
simultaneously explain the current data on the elastic proton form factor
and on Compton scattering in terms of perturbative QCD, without appealing
to large uncalculated higher-order and process-dependent corrections.


\vskip .4 cm 
\noindent
{\bf Acknowledgments}
\vskip .2 cm 

We thank Stan Brodsky for suggesting this work, and for useful discussions
and comments on the manuscript.  We are grateful to Marc Vanderhaeghen and
particularly Andreas Kronfeld for detailed discussions of their work and 
for providing us with unpublished data files.  We thank Markus Diehl for
several helpful remarks and suggestions.


\end{document}